\documentclass[prd,superscriptaddress,nofootinbib,amsfonts,notitlepage,longbibliography]{revtex4-1}
\usepackage[utf8]{inputenc}
\usepackage[T1]{fontenc}
\usepackage{hyperref}
\usepackage{graphicx,bm,amsmath,color,scalerel}
\usepackage{bbold}
\usepackage{wasysym}
\usepackage{verbatim}
\usepackage{amssymb}
\usepackage{epstopdf}
\usepackage{animate}
\usepackage{xmpmulti}
\usepackage{centernot}
\usepackage{multirow}
\usepackage{tikz}
\usepackage{comment}
\usepackage[normalem]{ulem} 
\usepackage{cancel}
\usepackage{empheq} 
\usepackage{feynmp-auto}
\usepackage{bigints}
\makeatletter

\newcounter{feynmancounter}
\newenvironment{feynman}[1]
  {
    \addtocounter{feynmancounter}{1}
    \begin{fmffile}{feynm\thefeynmancounter}
	\begin{fmfgraph*}(#1)
  }
  { 
    \end{fmfgraph*}
    \end{fmffile}
  }

\DeclareMathOperator{\Tr}{Tr}
\newcommand{\be}{\begin{equation}}
\newcommand{\ee}{\end{equation}}
\newcommand{\beq}{\begin{eqnarray}}
\newcommand{\eeq}{\end{eqnarray}}
\newcommand{\ba}{\begin{align}}
\newcommand{\ea}{\end{align}}

\begin{document}

\title{Decay of superluminal neutrinos in the collinear approximation}
\author{J.M. Carmona}
\email{jcarmona@unizar.es}
\affiliation{Departamento de F\'{\i}sica Te\'orica,
Universidad de Zaragoza, Zaragoza 50009, Spain}
\affiliation{Centro de Astropart\'{\i}culas y F\'{\i}sica de Altas Energ\'{\i}as (CAPA),
Universidad de Zaragoza, Zaragoza 50009, Spain}

\author{J.L. Cort\'es}
\email{cortes@unizar.es}
\affiliation{Departamento de F\'{\i}sica Te\'orica,
Universidad de Zaragoza, Zaragoza 50009, Spain}
\affiliation{Centro de Astropart\'{\i}culas y F\'{\i}sica de Altas Energ\'{\i}as (CAPA),
Universidad de Zaragoza, Zaragoza 50009, Spain}

\author{J.J. Relancio}
\email{jjrelancio@ubu.es}
\affiliation{Departamento de Matemáticas y Computación, Universidad de Burgos, 09001 Burgos, Spain}
\affiliation{Centro de Astropart\'{\i}culas y F\'{\i}sica de Altas Energ\'{\i}as (CAPA),
Universidad de Zaragoza, Zaragoza 50009, Spain}

\author{M.A. Reyes}
\email{mkreyes@unizar.es}
\affiliation{Departamento de F\'{\i}sica Te\'orica,
Universidad de Zaragoza, Zaragoza 50009, Spain}
\affiliation{Centro de Astropart\'{\i}culas y F\'{\i}sica de Altas Energ\'{\i}as (CAPA),
Universidad de Zaragoza, Zaragoza 50009, Spain}

\begin{abstract}
The kinematics of the three body decay, with a modified energy-momentum relation of the particles due to a violation of Lorentz invariance, is presented in detail in the collinear approximation. The results are applied to the decay of superluminal neutrinos producing an electron-positron or a neutrino-antineutrino pair. Explicit expressions for the energy distributions, required for a study of the cascade of neutrinos produced in the propagation of superluminal neutrinos, are derived.  
\end{abstract}

\maketitle

\section{Introduction}

In a theory of quantum gravity (QG), our classical notion of spacetime will surely be modified. It is natural to expect that the symmetries associated with the structure of spacetime (Poincaré invariance) will then also be modified. There are in fact indications from some candidates of QG that the Lorentz symmetry could be violated or deformed at very high energies~\cite{Colladay:1998fq,Amelino-Camelia2002b,Mattingly:2005re,AmelinoCamelia:2008qg,Liberati2013}. Lacking an understanding of the origin of the departure from Lorentz invariance, it is an open question how this departure will affect different particles.  Neutrinos are very especial ingredients of the Standard Model of particles physics (SM) due to their quantum numbers (behavior under the different interactions). This peculiarity may well be behind the origin of their extremely small masses and makes them very good candidates to look for physics beyond the SM, including a possible violation of the Lorentz symmetry. A manifestation of this violation is a modification of the standard relativistic expression for the energy as a function of the momentum, which, for Planckian (or another effective high-energy scale) corrections, will be more important at higher energies. In the case of modifications for which a higher value of the energy for a given momentum corresponds to a higher velocity (derivative of the energy with respect to the momentum) than in special relativity, neutrinos can become superluminal as a consequence of the Lorentz invariance violation (LIV).    

There has been an impressive progress in the  possibility of observing neutrinos of increasingly higher energies, including the recent observations of high-energy astrophysical neutrinos up to energies of the order of PeV~\cite{Aartsen:2013jdh,Aartsen:2014gkd,Aartsen:2018fqi}. There are prospects to extend this energy range by three orders of magnitude (EeV) in the near future (see the last section of Ref.~\cite{Stecker:2022tzd}). These observations would be affected by the modifications in the propagation of the neutrinos from their sources to their detection in the presence of LIV. In particular, superluminal neutrinos are no longer stable particles, being able to decay, producing an electron-positron pair or a neutrino-antineutrino pair, through the weak interaction. These decays would lead to a suppression, stronger at higher energies, of the detected flux of neutrinos.        

After the initial claim on superluminal propagation by the OPERA experiment in 2011~\cite{OPERA:2011ijqv1}, a number of theoretical models involving superluminal neutrinos were discussed in the literature~\cite{Alexandre:2011bu,Giudice:2011mm,Dvali:2011mn,Amelino-Camelia:2011mtr,Cacciapaglia:2011ax}.
Superluminal neutrinos can decay producing an electron-positron pair if the relation between the energy and momentum for electrons and positrons is not modified or if the modification is smaller than in the case of neutrinos. One can assume that this is the case, since there are very stringent limits on such modification of the energy-momentum relation for electrons (\cite{Jacobson:2002ye,Maccione:2007yc,Maccione:2011fr,Stecker:2013jfa,Rubtsov:2016bea,Altschul:2021wrm}, see additional references in~\cite{Addazi:2021xuf}); in fact, the inferred sensitivities for LIV in neutrinos that one would get from experiments involving the charged-lepton sector using gauge invariance arguments would exceed typical constraints in the neutrino sector by several orders of magnitude~\cite{Crivellin:2020oov}.

The investigation of phenomenological consequences of the decay of superluminal neutrinos involve the computation of decay rates in a LIV framework. This has mainly been studied in the case of new dimension 4 ($d=4$) operators in the free Lagrangian for the neutrino field, which gives a velocity of propagation for the neutrino which is different from $c$ and is energy-independent. Such a modification was used in the context of the former OPERA anomaly to give a first estimate of the production of an electron-positron pair by a superluminal neutrino by Cohen and Glashow~\cite{Cohen:2011hx}. Detailed calculations trying to reproduce the Cohen and Glashow result in different frameworks were given in Refs.~\cite{Bezrukov:2011qn,Huo:2011ve,Carmona:2012tp}. In particular, Ref.~\cite{Bezrukov:2011qn} showed that the result for the decay rate that is derived from a Lagrangian containing LIV terms is different from the Cohen and Glashow result, and depends on the explicit form of the interaction terms in the Lagrangian, computing the decay width for two different models (`model I' and `model II'~\cite{Bezrukov:2011qn}). These two models corresponded to the `second example' and `fourth example' of Ref.~\cite{Carmona:2012tp} (while the `first example' corresponded to the Cohen and Glashow result), where the dependence of decay rates on the choice of the dynamical matrix elements was also examined. Moreover, Ref.~\cite{Carmona:2012tp} considered different choices of modified dispersion relation for neutrinos, going beyond the case of an energy-independent velocity.

The choice of the `model II' in Ref.~\cite{Bezrukov:2011qn} and of the `fourth example' in Ref.~\cite{Carmona:2012tp} was motivated by a gauge invariance argument, which, as we will explain, is not satisfactory. There are in fact some concerns about the theoretical consistency of an scenario where all the effects of LIV are restricted to the neutrino sector. In the extension of the SM within the effective field theory framework~\cite{Colladay:1998fq}, one considers all possible terms involving the SM fields compatible with the gauge symmetry of the SM. This would lead to gauge covariant derivatives, instead of usual derivatives, acting on the $SU(2)$ doublet of left handed fields, including the neutrino and charged lepton fields. Then, one would have, in principle, together with the LIV corrections on neutrinos, a similar correction on the charged leptons, which may be incompatible with the above mentioned constraints in the charged lepton sector. It is a technical curiosity that, as explained in Ref.~\cite{Jentschura:2019wsr}, one can have different LIV parameters for a charged lepton and its neutrino in gauge-invariant models under a restricted set of gauge transformations, within the $SU(2)_L$ gauge group, if the models only involve the interaction with the $Z^0$ (so that the interaction Lagrangian is diagonal in $SU(2)_L$ space). The `model II' in Ref.~\cite{Bezrukov:2011qn} and the `fourth example' in Ref.~\cite{Carmona:2012tp} are precisely examples of this situation. Indeed, a limitation of all the above mentioned calculations of decay rates of superluminal neutrinos is that they were made considering only the neutral weak current. The complete model, however, contains the charged weak current, which means that the introduction of a LIV term at the level of the covariant derivatives is not a way to reconcile gauge invariance with a LIV affecting differently neutrinos and charged leptons.

There remain two possibilities to escape the argument that the LIV corrections should affect equally neutrinos as charged leptons. The first one is to assume that LIV corrections involve the lepton fields only through the gauge invariant product of the Higgs doublet and the lepton doublet~\cite{Carmona:2012tp}. Then, one can use derivatives of the invariant product and, when one replaces the Higgs doublet by its vacuum expectation value, the invariant product reduces to the neutrino field multiplied by a constant and one can obtain a LIV term involving only the neutrino field. This is one way to generate LIV effects affecting only the neutrino sector consistently with the gauge invariance of the SM. A more speculative alternative is the possibility that, together with the loss of Lorentz invariance, one had also to consider a departure from the gauge symmetries defining the Lorentz invariant SM. Indeed, as argued in~\cite{Jentschura:2020nfe}, LIV violates gauge invariance within general relativity. Lacking a well defined origin of the (possible) corrections to the SM, and also taking into account the very special role that neutrinos play within the SM, one should keep an open mind on the possibility of a relation between the violation of the Lorentz and the gauge symmetries of the SM, as previously pointed out in~\cite{Chkareuli:2007wh,Jentschura:2020nfe}. Any of these two possibilities leads to the introduction of LIV terms at the level of the free Lagrangian for the neutrino fields, and exclude these terms at the level of the interaction with the gauge fields. 

As indicated above, LIV effects motivated by quantum gravity are expected to become more relevant as the energy increases, which means a velocity of superluminal neutrinos which depends on the energy, or, more precisely, on $(E/\Lambda)^n$, where $\Lambda$ is the quantum-gravity-motivated LIV scale, and $n$ the order of the correction. The linear case, $n=1$, corresponds to $d=5$ operators in the Lagrangian, and the quadratic case, $n=2$, to $d=6$ operators. Besides this motivation, a correction due to LIV in the neutrino energy-momentum relation increasing with the energy provides a natural mechanism for the suppression of LIV effects at low energies, where one has the more precise tests of Lorentz invariance.

In~\cite{Mattingly:2009jf}, a first attempt to consider $n=2$ Planck-scale suppressed LIV effects on the cosmogenic neutrino spectrum was presented. An estimate of the decay width of a superluminal neutrino into three neutrinos (neutrino splitting), based on a rough approximation of the integral over the phase-space volume of the three particles in the final state, led to the prediction that one would have a cutoff at an energy in the interval $(10^{18} \,\text{eV}, 10^{19} \,\text{eV})$, preceded by a bump in the cosmogenic neutrino spectrum. Motivated by a hint of a suppression in the final part of the flux of astrophysical neutrinos detected by IceCube~\cite{IceCube:2013cdw}, a study of the possible effects of $n=1$ and $n=2$ LIV in the neutrino astrophysical spectrum at energies around the PeV scale was pursued in~\cite{Stecker:2014oxa,Carmona:2019xxp}, using the expressions of the decay width which had been obtained by the explicit calculations of Ref.~\cite{Carmona:2012tp}. The numerical results in Ref.~\cite{Stecker:2014oxa} contained, however, some uncertainties, because of two facts: the computation of Ref.~\cite{Carmona:2012tp} only included the pair-creation process, and only through the $Z^0$ exchange. Neutrino splitting had been the subject of a detailed calculation in Ref.~\cite{Somogyi:2019yis}, but only in the energy-independent (but flavor-dependent) velocity case.

The aim of this work is to go beyond the previous limitations and present a calculation that allows one to include both the neutrino splitting process and the charged weak bosons exchange in the computation of the decay width for a generic $n>0$ $(n\in \mathbb{N})$ neutrino superluminal case. We will do that by considering systematically the three body decay of a superluminal particle and will use this approach to determine the energy distribution of neutrinos in the decay of a superluminal neutrino. This may be useful for more detailed studies of the possible effects of this kind of LIV in the propagation of very high-energy superluminal neutrinos. The near future prospects to have a more precise determination of the neutrino astrophysical spectrum at energies above the PeV going up to EeV are a good motivation for such studies.

We will begin by briefly reviewing the introduction of superluminal neutrinos in the field theory framework in Sec.~\ref{sec:sl-nu}. As mentioned above, this will be done by including a LIV term of dimension $4+n$ in the free Lagrangian of the neutrino fields. In Sec.~\ref{sec:collinear}, we present in detail the collinear approximation to the three body decay of a superluminal particle, which is the main novelty of this work. This approximation is relevant when studying LIV corrections in a variety of situations in high-energy astrophysics, including the decay of a highly energetic particle. Indeed, in the following section we will apply the results of the collinear approximation to the decay of a superluminal neutrino producing an electron-positron pair (Sec.~\ref{sec:pp}) and a neutrino-antineutrino pair (Sec.~\ref{sec:spl}) for a LIV correction relevant to quantum gravity phenomenology ($n>0$). We present a summary of the results in Sec.~\ref{sec:conclusions}.

\section{Modified dispersion relation for superluminal neutrinos}
\label{sec:sl-nu}

We are going to consider the effects of LIV on the neutrino sector of the SM by adding to the SM Lagrangian  a LIV correction, compatible with rotational invariance, involving only the neutrino fields. In order to make this correction compatible with the very stringent limits on LIV, we assume that the LIV correction in the Lagrangian is a quadratic term in the neutrino fields with $(n+1)$ derivatives, so that its coefficient is the $n$-th power of the inverse of a new energy scale ($\Lambda$) parametrizing the LIV. Neutrino masses are irrelevant in the decays of superluminal neutrinos, which is the effect of the LIV correction we are going to study in this work, so we will treat them as massless particles. 

When treating the LIV as a first order correction to the Lorentz invariant SM Lagrangian, one can use the SM field equations to reduce the number of LIV terms. This means that in the terms quadratic in the neutrino fields one can replace any space derivative by a time derivative. As a consequence of the previous argument, one has a single LIV term in the Lagrangian
\be
{\cal L}_\text{LIV}^{(\nu)} \,=\, - \, \frac{1}{\Lambda^n} \, \overline{\nu}_{lL} \gamma^0 \, (i\partial_0)^{n+1} \, \nu_{lL}\,, 
\label{LIV-nu}
\ee
where the subindex $L$ refers to the left-handed chirality of the neutrino fields, $l$ refers to the three types of neutrinos ($e$, $\mu$, $\tau$), and we assume that there is no flavor dependence in the LIV terms, avoiding constrains from neutrino oscillations.
The choice of the sign in front of \eqref{LIV-nu} is arbitrary and will be discussed later.

The LIV term in the Lagrangian modifies the free theory of the neutrino field.  When we introduce a plane-wave expansion for the neutrino field 
\be
\nu_{L}(t, \vec{x}) \,=\, \int d^3\vec{k} \: \left[\tilde{b}_{\vec{k}} \, \tilde{u}(\vec{k}) \, e^{-i \tilde{E}_- t + i \vec{k}\cdot\vec{x}} + \tilde{d}^\dagger_{\vec{k}} \,\tilde{v}(\vec{k})\, e^{i \tilde{E}_+ t - i \vec{k}\cdot\vec{x}}\right]\,,
\ee
we find that the spinors $\tilde{u}$, $\tilde{v}$ have to satisfy the equations 
\be
\left[\gamma^0 \tilde{E}_- \,-\,\vec{\gamma}\cdot\vec{k} \,-\, \gamma^0 \frac{\tilde{E}_-^{n+1}}{\Lambda^n}\right] \,\tilde{u}(\vec{k}) \,=\, 0\,, \quad\quad
\left[\gamma^0 \tilde{E}_+ \,-\,\vec{\gamma}\cdot\vec{k} \,+\, (-1)^{n+1} \gamma^0 \frac{\tilde{E}_+^{n+1}}{\Lambda^n}\right] \,\tilde{v}(\vec{k}) \,=\, 0\,.
\ee
In the chiral representation for the Dirac matrices, the spinors $\tilde{u}$, $\tilde{v}$ can be written as $\tilde{u}=(\chi,0)^T$, $\tilde{v}=(\eta,0)^T$, and the bi-spinors $\chi$, $\eta$ satisfy the equations
\be
(\vec{\sigma}\cdot\vec{k}) \: \chi(\vec{k}) \,=\, - \left(\tilde{E}_- - \frac{\tilde{E}_-^{n+1}}{\Lambda^n}\right) \, \chi(\vec{k}) \,, \quad\quad\quad
(\vec{\sigma}\cdot\vec{k}) \: \eta(\vec{k}) \,=\, - \left(\tilde{E}_+ + (-1)^{n+1} \frac{\tilde{E}_+^{n+1}}{\Lambda^n}\right) \, \eta(\vec{k}) \,.
\ee
The matrix $(\vec{\sigma}\cdot\vec{k})$ has two eigenvalues $\pm |\vec{k}|$. Then we conclude that the relation between the momentum ($\vec{k}$) and the energy ($E_-$) for a neutrino is
\be
|\vec{k}| \,=\, \tilde{E}_- - \frac{\tilde{E}_-^{n+1}}{\Lambda^n}\,,
\label{mdr_neu}
\ee
and the relation between the momentum ($\vec{k}$) and the energy ($E_+$) for an antineutrino is
\be
|\vec{k}| \,=\, \tilde{E}_+ + (-1)^{n+1} \frac{\tilde{E}_+^{n+1}}{\Lambda^n}\,.
\label{mdr_anti}
\ee

We see then that, for the choice of minus sign in \eqref{LIV-nu}, when $n$ is even, both the neutrino and the antineutrino are superluminal, while in the case of $n$ odd the neutrino is superluminal and the antineutrino is subluminal. If we considered a positive coefficient in \eqref{LIV-nu} instead, any superluminal state would become subluminal and vice versa.

\section{Three-Body Decay Kinematics: collinear approximation}
\label{sec:collinear}

We present in this section a general analysis of the three body decay, $A(\vec{p}) \to B_1(\vec{p}_1) \,+\, B_2(\vec{p}_2) \,+\, B_3(\vec{p}_3)$, of a superluminal particle ($A$), with a momentum $\vec{p}$ and energy $E$, into three particles ($B_i$, $i=1,2,3$), with momentum $\vec{p}_i$ and energy $E_i$. We will consider the high-energy limit where one can neglect masses. 

We will assume an energy-momentum relation for each particle
\be
|\vec{p}| \approx E \,\left[1 - \left(\frac{E}{\Lambda}\right)^n\right]\,, \quad\quad\quad 
|\vec{p}_i| \approx E_i \,\left[1 + \alpha_i \left(\frac{E_i}{\Lambda}\right)^n\right]\,, 
\ee
where $\Lambda$ is the energy scale which parametrizes the LIV energy-momentum relation of the superluminal decaying particle ($A$). The dimensionless coefficients ($\alpha_i$) are introduced to consider the possibility to have different energy-momentum relations for the particles in the final state. We will consider two cases ($n=1,2$) for the exponent in the LIV (linear, quadratic) first contribution to the energy-momentum relation in an expansion in powers of $(E/\Lambda)$ and $(E_i/\Lambda)$. 

If one neglects terms proportional to the ratio of the energy and the scale of LIV in the energy-momentum relation of the particles, as in SR, the conservation of energy and momentum requires the momenta of the four particles to have the same direction in the high-energy limit. Then, considering a LIV scenario, the angles between the different momenta will be very small if $(E/\Lambda)\ll 1$. In fact, as we will check later, the square of the angle between any two particles is proportional to $(1/\Lambda)^n$. This is what we call collinear approximation, that will be used along the work.

We assume a (spin averaged) squared decay amplitude for a three-body decay given by~\footnote{We will show in the next sections that this expression applies to the squared amplitude of the weak decays of  superluminal neutrinos.}
\be
|{\cal A}|^2 \,=\, {\cal N} \,|\vec{p}| \,\vec{p}_1| \,\vec{p}_2| \,\vec{p}_3| \,(1-\widehat{p}\cdot\widehat{p}_1) \,(1-\widehat{p}_2\cdot\widehat{p}_3)\,,
\label{A2}
\ee
where ${\cal N}$ is a momentum independent constant and $\hat{v}\doteq (\vec{v}/|\vec{v}|)$ is an unitary vector. The factor $(1-\widehat p\cdot\widehat p_1) (1-\widehat p_2\cdot\widehat p_3)$ will be of order $O\left((1/\Lambda^n)^2\right)$; this fact will allow us to consider any other multiplicative factors involved in the calculation at dominant order only. Then, the final state three-body phase space can be approximated at dominant order by 
\be
\Phi_{\cal PS} \,\approx\, \left[\prod_{i=1}^3 \frac{d^3\vec{p}_i}{(2\pi)^3 \,2|\vec{p}_i|}\right] \,(2\pi)^4 \,\delta(E-\sum_i E_i) \,\delta^3(\vec{p}-\sum_i\vec{p}_i)\,.
\ee
The differential decay width will be given by
\be
d\Gamma\,=\,\frac{1}{2E}|\mathcal{A}|^2 \Phi_{\mathcal{PS}}\,,
\ee
and the full decay width will be
\be
\Gamma \,\approx\, \frac{{\cal N}}{16} \,\frac{1}{(2\pi)^5} \,\int \,\left[\prod_{i=1}^3 d^3\vec{p}_i\right] \,  \,\delta(E-\sum_i E_i) \,\delta^3(\vec{p}-\sum_i\vec{p}_i)\,
(1-\widehat{p}\cdot\widehat{p}_1) \,(1-\widehat{p}_2\cdot\widehat{p}_3)\,.
\label{Gamma(p)}
\ee

Prior to the integration over the three momenta, one can first identify the relation between the two angular dependent factors in the integrand of (\ref{Gamma(p)}). Those factors appear in the square of $(\vec{p}-\vec{p}_1)$ and $(\vec{p}_2+\vec{p}_3)$ respectively,
\be
(\vec{p}-\vec{p}_1)^2 \,=\, (|\vec{p}|-|\vec{p}_1|)^2 + 2 |\vec{p}| |\vec{p}_1| (1-\widehat{p}\cdot\widehat{p}_1), \quad\quad (\vec{p}_2+\vec{p}_3)^2 \,=\, (|\vec{p}_2|+|\vec{p}_3|)^2 - 2 |\vec{p}_2| |\vec{p}_3| (1-\widehat{p}_2\cdot\widehat{p}_3) \,.
\ee
Then, from the conservation of momentum, $\vec{p}-\vec{p}_1=\vec{p}_2+\vec{p}_3$, one gets
\begin{align}
& \hskip 3cm (1-\widehat{p}_2\cdot\widehat{p}_3) \,=\, \frac{(|\vec{p}_2|+|\vec{p}_3|)^2 - (|\vec{p}|-|\vec{p}_1|)^2}{2 |\vec{p}_2| |\vec{p}_3|} - \frac{|\vec{p}| |\vec{p}_1|}{|\vec{p}_2| |\vec{p}_3|} (1-\widehat{p}\cdot\widehat{p}_1) \nonumber \\
& \hskip 0.3cm \approx\, \frac{|\vec{p}|-|\vec{p}_1|}{|\vec{p}_2| |\vec{p}_3|} \,(|\vec{p}_2| + |\vec{p}_3| + |\vec{p}_1| - |\vec{p}|) - \frac{|\vec{p}| |\vec{p}_1|}{|\vec{p}_2| |\vec{p}_3|} (1-\widehat{p}\cdot\widehat{p}_1) \,\approx\, \frac{(1-x_1)}{x_2 x_3} \left[1 + \sum_i \alpha_i x_i^{n+1}\right] \left(\frac{E}{\Lambda}\right)^n - \frac{x_1}{x_2 x_3} (1-\widehat{p}\cdot\widehat{p}_1)\,. 
\label{theta23}
\end{align}
where in the last step we have introduced the energy fraction variables $x_i \doteq (E_i/E)$.

\begin{figure}[tb]
    \centering
    \includegraphics[width=0.5\textwidth]{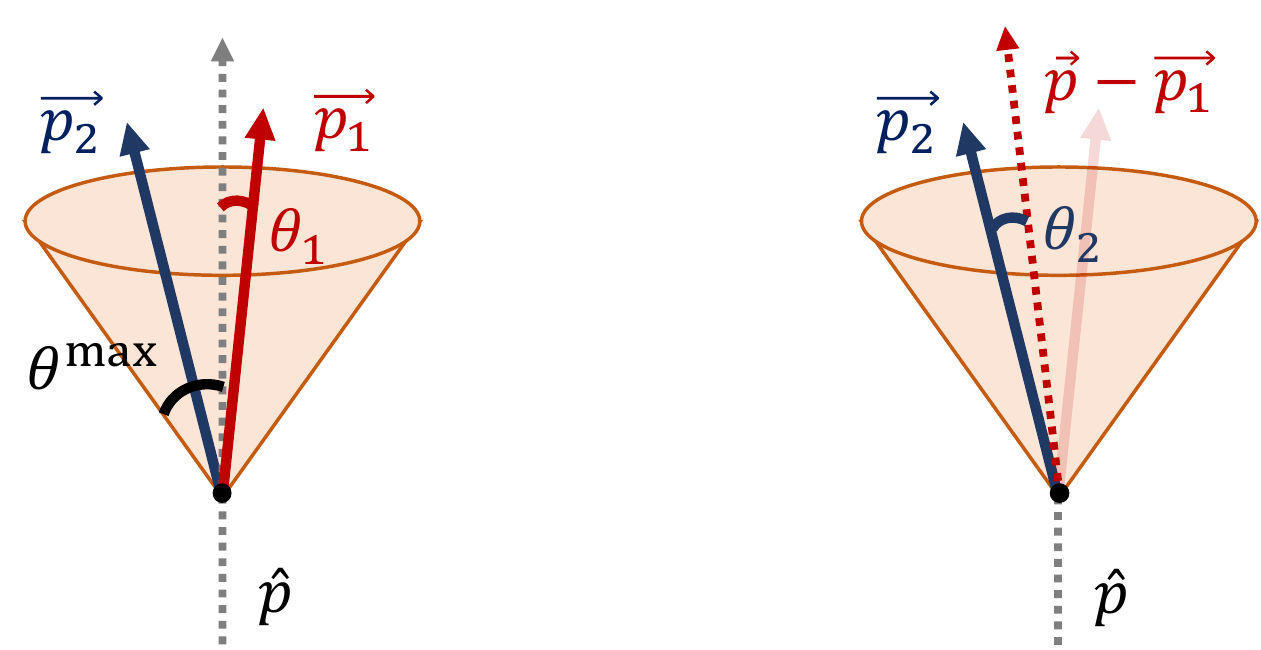}
    \caption{Cone of possible directions of the first and second particles of the final state. The direction of the second particle can be integrated in terms of the direction of the first one (right picture) and the direction of the first particle can be integrated using the spherical symmetry of the system (left picture).}
    \label{fig:disin_cone}
\end{figure}

Additionally, the conservation of momentum can be used to integrate one of the three momentum variables, for example, $\vec{p}_3$. The other two momenta ($\vec p_1, \vec p_2$) should be integrated within a cone of apperture $\theta^\text{max}$, as shown in Fig.~\ref{fig:disin_cone}.
The integral over $\vec p_2$ can be done as a function of the direction of $\vec{p}_1$. For that, let us use spherical coordinates whose $z$ axis is given by the direction of the vector $(\vec{p}-\vec{p}_1)$ (Fig.~\ref{fig:disin_cone}, right). In order to cover all the cone, one can make the azimuthal angle to go from 0 to $2\pi$, which gives a factor $2\pi$, while $\theta_2$, the angle between the two vectors $(\vec{p}-\vec{p}_1)$ and $\vec{p}_2$, must go between two functions of $\vec{p}_1$. Even if these functions could be complicated, one can perform the integral over $\theta_2$ using the energy Dirac delta in the following way. From the expression of the square of the momentum $\vec{p}_3$, one gets
\be
\vec{p}_3^{\,2} = (\vec{p} - \vec{p}_1 - \vec{p}_2)^2 = |\vec{p}-\vec{p}_1|^2 + |\vec{p}_2|^2 - 2 |\vec{p}-\vec{p}_1| |\vec{p}_2| \cos\theta_2\,;
\ee
then, as the direction of $\vec p$ and $\vec p_1$ are fixed, one can relate the angle $\theta_2$ with the energy fraction $x_3$ as 
\be
d(\cos\theta_2) \,=\, - \frac{|\vec{p}_3|}{|\vec{p}-\vec{p}_1| |\vec{p}_2|} \,d|\vec{p}_3| \,\approx\, - \frac{x_3}{(1-x_1) x_2}\, dx_3\,.
\ee
This way, the integral over the angular variable $\theta_2$ can expressed as an integral over $x_3$, where we can use the energy delta to integrate. Now, we have to perform the integration over $\vec p_1$. For that, let us use spherical coordinates whose $z$ axis is given by the direction of the parent particle, $\vec p$ (Fig.~\ref{fig:disin_cone}, left). Then the azimuthal angle goes from 0 to $2\pi$, which gives again a factor $2\pi$, and the angular variable $\theta_1$ from 0 to $\theta^\text{max}$. Defining $\omega_1 \doteq (1-\cos\theta_1)$, which goes from 0 to $\omega^\text{max}$, one has 
\be
\Gamma \,\approx\, \frac{{\cal N}}{16} \,\frac{E^5}{(2\pi)^3}\,\int dx_1 \,dx_2\,dx_3\,\delta(1-\sum_i x_i)\,d\omega_1\,\frac{x_1^2 x_2 x_3}{(1-x_1)}\,\left\{\frac{(1-x_1)}{x_2 x_3} \left[1 + \sum_i \alpha_i x_i^{n+1}\right] \left(\frac{E}{\Lambda}\right)^n \omega_1 - \frac{x_1}{x_2 x_3} \omega_1^2\right\}\,,
\ee
where $\omega^\text{max}$ can be read from (\ref{theta23}) to be 
\be
\omega^\text{max} \,\approx\, \frac{(1-x_1)}{x_1} \,(1 + \sum_i \alpha_i x_i^{n+1})  \left(\frac{E}{\Lambda}\right)^n\,.
\ee

Then, the final expression for the decay width as an integral over the energy fractions $x_i$ is
\be
\Gamma \,\approx\, \frac{{\cal N}}{96} \,\frac{E^5}{(2\pi)^3}\,\left(\frac{E}{\Lambda}\right)^{3n}\, \bigintsss \left[\prod_i dx_i\right]\,\delta(1-\sum_i x_i)\,(1-x_1)^2\,\left[1 + \sum_i \alpha_i x_i^{n+1}\right]^3\,,
\label{Gamma(x)}
\ee
from which one can read the differential decay width, 
\be
\frac{d^3\Gamma}{dx_1\,dx_2\,dx_3}\, =\, \frac{{\cal N}}{96} \,\frac{E^5}{(2\pi)^3}\,\left(\frac{E}{\Lambda}\right)^{3n}\,\delta(1-\sum_i x_i)\,(1-x_1)^2\,\left[1 + \sum_i \alpha_i x_i^{n+1}\right]^3\,\,,
\label{difGamma}
\ee
and the energy distribution after the decay
\be
{\cal P}(x_1, x_2, x_3) \,\doteq\, \frac{1}{\Gamma} \,\frac{d^3\Gamma}{dx_1\,dx_2\,dx_3}\,.
\ee

\section{Application of the collinear approximation to the decay of superluminal neutrinos}

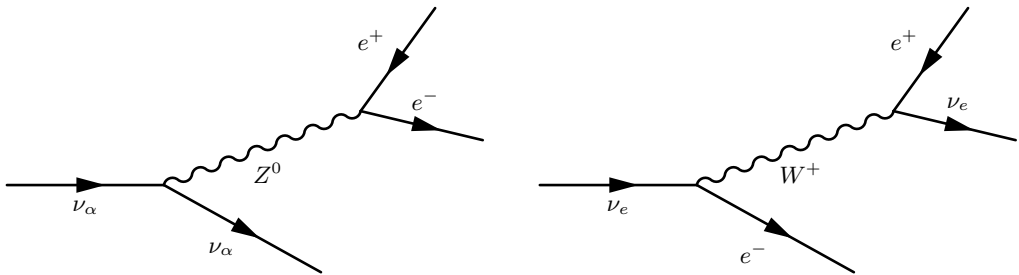
\begin{figure}[tb]
    \centering
    \begin{feynman}{180,100}
        \fmfleft{i1}
        \fmfright{o1,o2,o3}
        \fmf{fermion,label=$\nu_\alpha$}{i1,v1}
        \fmf{fermion,label=$\nu_\alpha$}{v1,o1}
        \fmf{boson,label=$Z^0$}{v1,v2}
        \fmf{fermion,label=$e^-$}{v2,o2}
        \fmf{fermion,label=$e^+$}{o3,v2}
        \fmfforce{(0.00w,0.33h)}{i1}
        \fmfforce{(0.33w,0.33h)}{v1}
        \fmfforce{(0.66w,0.00h)}{o1}
    \end{feynman}
    \begin{feynman}{180,100}
        \fmfleft{i1}
        \fmfright{o1,o2,o3}
        \fmf{fermion,label=$\nu_e$}{i1,v1}
        \fmf{fermion,label=$e^-$}{v1,o1}
        \fmf{boson,label=$W^+$}{v1,v2}
        \fmf{fermion,label=$\nu_e$}{v2,o2}
        \fmf{fermion,label=$e^+$}{o3,v2}
        \fmfforce{(0.00w,0.33h)}{i1}
        \fmfforce{(0.33w,0.33h)}{v1}
        \fmfforce{(0.66w,0.00h)}{o1}
    \end{feynman}
    \caption{Disintegration of a neutrino into a pair electron-positron and another neutrino, mediated by a $Z^0$ boson (left) or by a $W^+$ boson (right).}
    \label{fig:feynman_vpe}
\end{figure}

As explained in the introduction, in order to have LIV effects affecting neutrinos without the strong constraints imposed to LIV in the charged lepton sector, we need to assume that LIV corrections appear only at the level of the free neutrino Lagrangian, Eq.~\eqref{LIV-nu}, so that the weak interaction Lagrangian we will use to determine the decay widths of superluminal neutrinos will be the same as in the SR theory,
\be
{\cal L}_{\text{int}} \,=\, - \frac{g}{\sqrt{2}} \left[W_\mu^- \, \overline{e}_L \gamma^\mu \nu_{eL} + W_\mu^+ \, \overline{\nu}_{eL} \gamma^\mu e_L\right] - \frac{g}{2 c_W} Z_\mu \, \overline{\nu}_{lL} \gamma^\mu \nu_{lL} - \frac{g}{c_W} (s_W^2 - 1/2) Z_\mu \, \overline{e}_L \gamma^\mu e_L - \frac{g}{c_W} s_W^2 Z_\mu \, \overline{e}_R \gamma^\mu e_R\,,
\ee
which now includes the emission of electron-positron pairs through neutral (Fig.~\ref{fig:feynman_vpe}, left) and charged (Fig.~\ref{fig:feynman_vpe}, right) channels, and the disintegration through neutrino splitting (Fig.~\ref{fig:feynman_NSpl}). The first process, vacuum pair emission, has an energy threshold of~\cite{Carmona:2012tp} $E_{\text{th}} = (2m_e^2\Lambda^n)^{1/(2+n)}$ (it only occurs for neutrino energies exceding this value, $E>E_{\text{th}}$), while the second process, neutrino splitting, has a negligible threshold (zero in the limit of massless neutrinos).

\subsection{Electron-positron pair production}
\label{sec:pp}

We can start with the simplest case, which is the production of an electron-positron pair by a $\nu_\mu$ or $\nu_\tau$, since this process only involves the neutral current interaction. The amplitudes for the production of an electron-positron pair by a $\nu_\mu$ or $\nu_\tau$ are
\begin{align}
&{\cal A}(\nu_{\mu,\tau}(\vec{p})\to\nu_{\mu,\tau}(\vec{p'}) e_L^-(\vec{p}_-) e_L^+(\vec{p}_+))\,=\, \left(-i \frac{g}{2 c_W}\right) \, \left(-i \frac{g}{c_W}\right) (s_W^2-1/2) \, \frac{-i g_{\mu\nu}}{M_Z^2} \, [\bar{\tilde{u}}(\vec{p'}) \gamma^\mu \tilde{u}(\vec{p})] \, [\bar{u}_L(\vec{p}_-) \gamma^\nu v_L(\vec{p}_+)]\,, \nonumber \\
&{\cal A}(\nu_{\mu,\tau}(\vec{p})\to\nu_{\mu,\tau}(\vec{p'}) e_R^-(\vec{p}_-) e_R^+(\vec{p}_+))\,=\, \left(-i \frac{g}{2 c_W}\right) \, \left(-i \frac{g}{c_W}\right) s_W^2 \, \frac{-i g_{\mu\nu}}{M_Z^2} \, [\bar{\tilde{u}}(\vec{p'}) \gamma^\mu \tilde{u}(\vec{p})] \, [\bar{u}_R(\vec{p}_-) \gamma^\nu v_R(\vec{p}_+)]\,,
\label{A-nu(mutau)-e+e-}
\end{align}
where ($u_L$, $u_R$) are the spinors describing the two possible spin states of the electron, ($v_L$, $v_R$) the spinors for the two spin states of the positron, and $\tilde{u}$ the spinor of the only state of the superluminal neutrino. Let us notice that, as pointed out in~\cite{Somogyi:2019yis,Jentschura:2020nfe}, some works like~\cite{Cohen:2011hx,Huo:2011ve,Carmona:2012tp} considered two active spin states for the neutrino, while others, like~\cite{Bezrukov:2011qn,Jentschura:2020nfe}, assume a single possible spin state. This assumption affects the average over the initial states and the final result of the decay width. In the present work, we take the neutrino as a massless particle, and so, with a given helicity. This is a good approximation in the high-energy limit where all the particle momenta have approximately the same direction (collinear approximation). The use of this approximation is also what allowed us to neglect the four-momentum $q^2\doteq(p-p_1)^2\ll M_Z$ in the $Z^0$ propagator. 
Then, one can use the identities
\be
u_L(\vec{k}) \bar{u}_L(\vec{k}) \,=\, \tilde{u}(\vec{k}) \bar{\tilde{u}}(\vec{k}) \,=\, \left(\frac{1-\gamma_5}{2}\right) \cancel{k} \,, \quad\quad 
u_R(\vec{k}) \bar{u}_R(\vec{k}) \,=\, \left(\frac{1+\gamma_5}{2}\right) \cancel{k} \,,
\label{u-ubar}
\ee
where $\cancel{k}=k_\mu \gamma^\mu$ with $k_0\doteq |\vec{k}|$, to write the squared amplitudes as a product of traces
\begin{align}
&|{\cal A}(\nu_{\mu,\tau}(\vec{p})\to\nu_{\mu,\tau}(\vec{p'}) e_L^-(\vec{p}_-) e_L^+(\vec{p}_+))|^2 \,=\, \left(\frac{g^2}{M_W^2}\right)^2 (s_W^2-1/2)^2 \, g_{\mu\nu} \, g_{\rho\sigma} \nonumber \\
&\quad\quad\quad \times\frac{1}{4} \, \Tr\left[\left(\frac{1-\gamma_5}{2}\right) \cancel{p} \gamma^\rho \left(\frac{1-\gamma_5}{2}\right) \cancel{p}\,' \gamma^\mu\right] \quad \Tr\left[\left(\frac{1-\gamma_5}{2}\right) \cancel{p}_+ \gamma^\sigma \left(\frac{1-\gamma_5}{2}\right) \cancel{p}_- \gamma^\nu\right] \,, \nonumber \\
&|{\cal A}(\nu_{\mu,\tau}(\vec{p})\to\nu_{\mu,\tau}(\vec{p'}) e_R^-(\vec{p}_-) e_R^+(\vec{p}_+))|^2 \,=\, \left(\frac{g^2}{M_W^2}\right)^2 (s_W^2)^2 \, g_{\mu\nu} \, g_{\rho\sigma} \nonumber \\
&\quad\quad\quad \times \frac{1}{4} \, \Tr\left[\left(\frac{1-\gamma_5}{2}\right) \cancel{p} \gamma^\rho \left(\frac{1-\gamma_5}{2}\right) \cancel{p}\,' \gamma^\mu\right] \quad \Tr\left[\left(\frac{1+\gamma_5}{2}\right) \cancel{p}_+ \gamma^\sigma \left(\frac{1+\gamma_5}{2}\right) \cancel{p}_- \gamma^\nu\right]\,.
\end{align}
Using trace theorems and properties of $\gamma$-matrices, one has 
\begin{align}
&|{\cal A}(\nu_{\mu,\tau}(\vec{p})\to\nu_{\mu,\tau}(\vec{p'}) e_L^-(\vec{p}_-) e_L^+(\vec{p}_+))|^2 \,=\, \left(\frac{g^2}{M_W^2}\right)^2 (s_W^2-1/2)^2 \, 4 \, (p\cdot p_+) (p'\cdot p_-) \,, \label{PP-L} \\
&|{\cal A}(\nu_{\mu,\tau}(\vec{p})\to\nu_{\mu,\tau}(\vec{p'}) e_R^-(\vec{p}_-) e_R^+(\vec{p}_+))|^2 \,=\, \left(\frac{g^2}{M_W^2}\right)^2 (s_W^2)^2 \,
4 \, (p\cdot p_-) (p'\cdot p_+) \,. 
\label{PP-R}
\end{align}
Taking into account that $k_0\doteq|\vec{k}|$, we have 
\begin{align}
& (p\cdot p_+) (p'\cdot p_-) \,=\, |\vec{p}| \, |\vec{p'}| \, |\vec{p}_-| \, |\vec{p}_+| \, (1 - \hat{p}\cdot \hat{p}_+) \, (1 - \hat{p}\,'\cdot \hat{p}_-) \,, \nonumber \\
& (p\cdot p_-) (p'\cdot p_+) \,=\, |\vec{p}| \, |\vec{p'}| \, |\vec{p}_-| \, |\vec{p}_+| \, (1 - \hat{p}\cdot \hat{p}_-) \, (1 - \hat{p}\,'\cdot \hat{p}_+) \,.
\end{align}
As mentioned in the previous section, one can check that, when the directions of the four momenta coincide, the factors depending on the directions of the momenta and the squared amplitudes are zero. Due to the LIV correction, the energy-momentum conservation relations allow for a difference in the directions of the momenta of the particles and then one has a non-null result for the squared amplitudes.

The squared amplitudes in \eqref{PP-L} and \eqref{PP-R} have the general form (\ref{A2}) used in the previous section. In the case of the decay $\nu_{\mu,\tau}(\vec{p})\to\nu_{\mu,\tau}(\vec{p'}) e_L^-(\vec{p}_-) e_L^+(\vec{p}_+)$, one has 
\be
{\cal N} \,=\, 4\, \left(\frac{g^2}{M_W^2}\right)^2 \,(s_W^2 - 1/2)^2\,,
\ee
and $\vec{p}_1 = \vec{p}_+$, $\vec{p}_2 = \vec{p'}$, $\vec{p}_3 = \vec{p}_-$. The energy-momentum relation for the electron and the positron is not modified (we are considering a LIV contribution only in the energy-momentum relation of neutrinos) so that $\alpha_2=-1$, $\alpha_1=\alpha_3=0$. The decay width of a $\nu_\mu$ or $\nu_\tau$ producing an electron-positron pair can then be read from Eq.~\eqref{Gamma(x)},
\be
\Gamma(\nu_{\mu,\tau}(E)\to \nu_{\mu,\tau}\,e_L^-\,e_L^+) \,\approx\, \left(\frac{g^2}{M_W^2}\right)^2 \,(s_W^2 - 1/2)^2 \,\frac{E^5}{192\,\pi^3} \,\left(\frac{E}{\Lambda}\right)^{3n} \,\int dx' \,dx_- \,dx_+ \,\delta(1-x'-x_--x_+) (1-x'^{\,n+1})^3 (1-x_+)^2 
\,.
\label{Gamma(PP-L)}
\ee

In the case of the decay $\nu_{\mu,\tau}(\vec{p})\to\nu_{\mu,\tau}(\vec{p'}) e_R^-(\vec{p}_-) e_R^+(\vec{p}_+)$, one has 
\be
{\cal N} \,=\, 4\, \left(\frac{g^2}{M_W^2}\right)^2 \,(s_W^2)^2 \,,
\ee
and $\vec{p}_1 = \vec{p}_-$, $\vec{p}_2 = \vec{p'}$, $\vec{p}_3 = \vec{p}_+$. Then 
\be
\Gamma(\nu_{\mu,\tau}(E)\to \nu_{\mu,\tau}\,e_R^-\,e_R^+) \,\approx\, \left(\frac{g^2}{M_W^2}\right)^2 \,(s_W^2)^2 \,\frac{E^5}{192\,\pi^3} \,\left(\frac{E}{\Lambda}\right)^{3n} \,\int dx' \,dx_- \,dx_+ \,\delta(1-x'-x_--x_+) (1-x'^{\,n+1})^3 (1-x_-)^2 
\,.
\label{Gamma(PP-R)}
\ee
Combining the results in (\ref{Gamma(PP-L)}) and (\ref{Gamma(PP-R)}), we have that the decay width of a $\nu_\mu$ or $\nu_\tau$ producing an electron-positron pair is
\begin{align}
\Gamma(\nu_{\mu,\tau}(E)\to \nu_{\mu,\tau} \,e^-\,e^+) \,\approx\, & \left(\frac{g^2}{M_W^2}\right)^2 \,\frac{E^5}{192\,\pi^3} \,\left(\frac{E}{\Lambda}\right)^{3n} \,\int dx' \,dx_- \,dx_+ \,\delta(1-x'-x_--x_+) \nonumber \\ & \times (1-x'^{\,n+1})^3 \left[(s_W^2 - 1/2)^2 (1-x_+)^2 + (s_W^2)^2 (1-x_-)^2\right] \,.
\label{Gamma(PP)}
\end{align}
If we integrate the expression in \eqref{Gamma(PP)} over $x_+$, $x_-$ we obtain 
\be
\Gamma(\nu_{\mu,\tau}(E)\to \nu_{\mu,\tau} \,e^-\,e^+) \,\approx\, \left(\frac{g^2}{M_W^2}\right)^2 \,\frac{E^5}{192\,\pi^3} \left(\frac{E}{\Lambda}\right)^{3n} \, \left[\left(s_W^2-\frac{1}{2}\right)^2 + \left(s_W^2\right)^2\right]\,\int dx' \, \frac{1}{3} \,(1-x'^{\,n+1})^3 \,(1-x'^3)\,.
\ee
Then, the total decay width of production of an electron-positron pair is
\begin{empheq}[box=\boxed]{equation}
\Gamma(\nu_{\mu,\tau}(E)\to \nu_{\mu,\tau} \,e^-\,e^+) \,\approx\, \left(\frac{g^2}{M_W^2}\right)^2 \,\frac{E^5}{192\,\pi^3} \left(\frac{E}{\Lambda}\right)^{3n} \, \left[\left(s_W^2-\frac{1}{2}\right)^2 + \left(s_W^2\right)^2\right] \,c_n^{(e)}
\,,
\label{Gamma(nux-ee)}
\end{empheq}
where 
\be
c_n^{(e)} \,=\, \left[\frac{1}{4} - \frac{3}{(n+2)(n+5)} + \frac{3}{(2n+3)(2n+6)} - \frac{1}{(3n+4)(3n+7)}\right]\,.
\ee
The energy distribution is given by
\begin{empheq}[box=\boxed]{align}
{\cal P}_{\nu_{\mu,\tau} e^- e^+/\nu_{\mu,\tau}} (x',x_-,x_+) \,=\, &\frac{1}{\,c_n^{(e)}} \,\delta(1-x'-x_--x_+)\,(1-x'^{\,n+1})^3 \notag \\ & \times \left[\frac{(s_W^2 - 1/2)^2}{(s_W^2 - 1/2)^2+(s_W^2)^2} (1-x_+)^2 + \frac{(s_W^2)^2}{(s_W^2 - 1/2)^2+(s_W^2)^2} (1-x_-)^2\right]\,.
\label{P-mu-tau}
\end{empheq}
If we focus on the energy loss distribution of the neutrino, we get
\be 
{\cal P}_{\nu_{\mu,\tau}} (x) \,\doteq\, \int dx_-\, dx_+ \, {\cal P}_{\nu_{\mu,\tau} e^- e^+/\nu_{\mu,\tau}} (x,x_-,x_+) \,=\,\frac{1}{3\,c_n^{(e)}} \,(1-x^{n+1})^3 \,(1-x^3)\,.
\ee

Let us now consider the production of an electron-positron pair in the decay of a $\nu_e$. The result for the amplitude of the production of an ($e_L^-e_L^+$) pair has an additional contribution due to the charged current interaction,
\begin{align}
{\cal A}(\nu_e(\vec{p})\to\nu_e(\vec{p'}) \,e_L^-(\vec{p}_-) \, e_L^+(\vec{p}_+)) \,=& \left(-i \frac{g}{2 c_W}\right) \, \left(-i \frac{g}{c_W}\right) (s_W^2-1/2) \, \frac{-i g_{\mu\nu}}{M_Z^2} \, [\bar{\tilde{u}}(\vec{p'}) \gamma^\mu \tilde{u}(\vec{p})] \, [\bar{u}_L(\vec{p}_-) \gamma^\nu v_L(\vec{p}_+)]  \nonumber \\
& +\left(-i \frac{g}{\sqrt{2}}\right)^2 \,  \frac{-i g_{\mu\nu}}{M_W^2} \, [\bar{\tilde{u}}(\vec{p'}) \gamma^\mu v_L(\vec{p}_+)] \, [\bar{u}_L(\vec{p}_-) \gamma^\nu \tilde{u}(\vec{p})]\,.
\end{align}
Using trace theorems and properties of $\gamma$-matrices, one has 
\be
|{\cal A}(\nu_e\to\nu_e \,e_L^- \,e_L^+)|^2 \,=\, \left(\frac{g^2}{M_W^2}\right)^2 \, \left[(s_W^2-1/2)^2 - 2 (s_W^2-1/2) +1\right] \,4 (p\cdot p_+) (p'\cdot p_-)\,,     
\ee
which is just the result for the $\nu_{\mu,\tau}$ decay replacing $(s_W^2 - 1/2)^2$ by $(s_W^2 - 3/2)^2$. There is no contribution due to the charged current interaction in the production of an ($e_R^-e_R^+$) pair, and the amplitude for this decay is the same as in the decay of a $\nu_{\mu,\tau}$. Then, the decay width of a $\nu_e$ producing an electron-positron pair is
\begin{align}
\Gamma(\nu_e(E)\to \nu_e \,e^-\,e^+) \,\approx\, & \left(\frac{g^2}{M_W^2}\right)^2 \,\frac{E^5}{192\,\pi^3} \,\left(\frac{E}{\Lambda}\right)^{3n} \,\int dx' \,dx_- \,dx_+ \,\delta(1-x'-x_--x_+) \nonumber \\ & \times (1-x'^{\,n+1})^3 \left[(s_W^2 - 3/2)^2 (1-x_+)^2 + (s_W^2)^2 (1-x_-)^2\right] \,,
\label{Gamma(PP)nue}
\end{align}
instead of (\ref{Gamma(PP)}). Integrating, one finds that the total decay width for a $\nu_e$ producing an electron-positron pair is
\begin{empheq}[box=\boxed]{equation}
\Gamma(\nu_e(E)\to \nu_e \,e^-\,e^+) \,\approx\, \left(\frac{g^2}{M_W^2}\right)^2 \,\frac{E^5}{192\,\pi^3} \left(\frac{E}{\Lambda}\right)^{3n} \,  \left[\left(s_W^2-\frac{3}{2}\right)^2 + \left(s_W^2\right)^2\right] \,c_n^{(e)}
\,,
\label{Gamma(nue-ee)}
\end{empheq}
and the corresponding energy distribution is given by
\begin{empheq}[box=\boxed]{align}
{\cal P}_{\nu_e e^- e^+/\nu_e} (x',x_-,x_+) \,=\, &\frac{1}{\,c_n^{(e)}} \,\delta(1-x'-x_--x_+)\,(1-x'^{\,n+1})^3 \notag \\ & \times\left[\frac{(s_W^2 - 3/2)^2}{(s_W^2 - 3/2)^2+(s_W^2)^2} (1-x_+)^2 + \frac{(s_W^2)^2}{(s_W^2 - 3/2)^2+(s_W^2)^2} (1-x_-)^2\right]\,.
\label{P-e}
\end{empheq}

Let us notice that the total decay widths of $\nu_e$ and of $\nu_{\mu,\tau}$ differ only by a constant factor. If we focus only on the energy loss of the neutrino, we find a flavor independent distribution,
\begin{empheq}[box=\boxed]{equation}
{\cal P}_{\nu_e} (x) \,=\,{\cal P}_{\nu_\mu} (x) \,=\, {\cal P}_{\nu_\tau} (x) \,=\,\frac{1}{3\,c_n^{(e)}} \,(1-x^{n+1})^3 \,(1-x^3)\,.
\label{nu-nu-PP}
\end{empheq}

There are two special cases, linear or quadratic LIV modification of the energy-momentum relation, corresponding to $n=1$ or $n=2$. The total decay widths and neutrino energy loss distribution for the production of an electron-positron pair are obtained by replacing the constant $c_n^{(e)}$ in (\ref{Gamma(nux-ee)}), (\ref{Gamma(nue-ee)}), and (\ref{nu-nu-PP}),  by 
\be
c_1^{(e)} \,=\, \frac{121}{840}\approx 0.144\,, \hskip 2cm c_2^{(e)} \,=\, \frac{81}{455}\approx 0.178\,.
\label{c(e)number}
\ee
Another difference between the two cases is that, while in the linear case the antineutrinos are subluminal (and then they are stable) when the neutrinos are superluminal, in the quadratic case both neutrinos and antineutrinos are superluminal, allowing the antineutrino to also decay. In order to obtain the final energy distribution of the antineutrino electron-positron pair emission, one should take equations \eqref{P-mu-tau} and \eqref{P-e} and exchange every particle by its antiparticle, i.e., the energy distribution of the electron and positron should be swapped. The expressions for the total decay width and energy loss apply to both neutrinos and antineutrinos. 

The total decay width, as well as the energy loss distribution of a superluminal neutrino producing an electron-positron pair, was obtained in Ref.~\cite{Carmona:2012tp} by a different method, and considering only the $Z^0$ exchange. The calculation of the $\nu_{\mu,\tau}$ decay made here corresponds then to the second example examined in that work, which was derived from a modified Lagrangian, as we do in the present work, for the case of a $\Lambda$-suppressed modification of the energy-momentum relation for the neutrino. 
One can check that the calculation based on the collinear approximation reproduces the result obtained in Ref.~\cite{Carmona:2012tp} for the neutrino energy loss distribution and also the result for the total decay width, up to a factor of two (compare, e.g., Eq.~\eqref{Gamma(nux-ee)} with Eq.~(52) of Ref.~\cite{Carmona:2012tp}). The origin of the discrepancy is that in~\cite{Carmona:2012tp}, as well as in~\cite{Cohen:2011hx}, one makes an average over two possible initial neutrino spin states. However, as discussed previously, the two initial states are not equally probable, and can be approximated by only one helicity state of a massless particle. 

\subsection{Neutrino-antineutrino pair production}
\label{sec:spl}

\begin{figure}[tb]
    \centering
    \begin{feynman}{180,100}
        \fmfleft{i1}
        \fmfright{o1,o2,o3}
        \fmf{fermion,label=$\nu_\alpha$}{i1,v1}
        \fmf{fermion,label=$\nu_\alpha$}{v1,o1}
        \fmf{boson,label=$Z^0$}{v1,v2}
        \fmf{fermion,label=$\nu_\beta$}{v2,o2}
        \fmf{fermion,label=$\bar\nu_\beta$}{o3,v2}
        \fmfforce{(0.00w,0.33h)}{i1}
        \fmfforce{(0.33w,0.33h)}{v1}
        \fmfforce{(0.66w,0.00h)}{o1}
    \end{feynman}
    \caption{Disintegration of a neutrino in two neutrinos and one antineutrino, mediated by a $Z^0$ boson.}
    \label{fig:feynman_NSpl}
\end{figure}
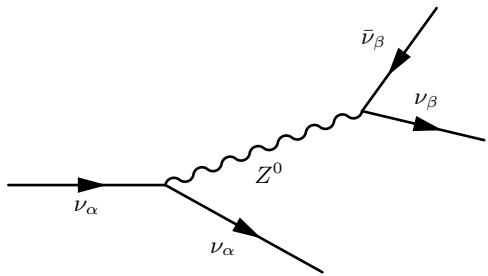

We consider a neutrino-antineutrino pair of flavor $\beta$ produced in the decay of a neutrino of a different flavor $\alpha$ (Fig.~\ref{fig:feynman_NSpl}).
The amplitude of the decay is
\be
{\cal A}(\nu_\alpha(\vec{p}) \to \nu_\alpha(\vec{p'}) \,\nu_\beta(\vec{p}_-)\,\bar{\nu}_\beta(\vec{p}_+)) \,=\, \left(-i \frac{g}{2c_W}\right)^2 \,\frac{-i g_{\mu\nu}}{M_Z^2} \left[\bar{\tilde{u}}(\vec{p'}) \gamma^\mu \tilde{u}(\vec{p})\right] \,
\left[\bar{\tilde{u}}(\vec{p}_-) \gamma^\nu \tilde{v}(\vec{p}_+)\right]\,,
\ee
and the squared amplitude
\begin{align}
|{\cal A}|^2 \,=& \left(\frac{g^2}{M_W^2}\right)^2 \, g_{\mu\nu} g_{\rho\sigma} \,\frac{1}{16} \, \Tr\left[\left(\frac{1-\gamma_5}{2}\right) \cancel{p} \, \gamma^\rho \left(\frac{1-\gamma_5}{2}\right) \cancel{p}\,' \, \gamma^\mu\right] \Tr\left[\left(\frac{1-\gamma_5}{2}\right) \cancel{p}_+ \, \gamma^\sigma \left(\frac{1-\gamma_5}{2}\right) \cancel{p}_- \, \gamma^\nu\right] \nonumber \\ \,=& \left(\frac{g^2}{M_W^2}\right)^2 \, (p\cdot p_+) (p'\cdot p_-)\,. 
\end{align}
This is once more the expression~(\ref{A2}) used in Sec.~\ref{sec:collinear} with 
\be
{\cal N} \,=\, \left(\frac{g^2}{M_W^2}\right)^2\,,
\ee
and $\vec{p}_1 = \vec{p}_+$, $\vec{p}_2 = \vec{p'}$, $\vec{p}_3 = \vec{p}_-$. The energy-momentum relation for the neutrinos in the final state has a LIV correction coefficient $\alpha_2 =\alpha_3=-1$, while the coefficient in the antineutrino energy-momentum relation is $\alpha_1^{(1)}=1$ in the linear case ($n=1$), or $\alpha_1^{(2)}=-1$ in the quadratic one ($n=2$). 
Using Eq.~(\ref{Gamma(x)}), the decay width of a neutrino-antineutrino pair production is
\begin{align}
\Gamma(\nu_\alpha(E)\to \nu_\alpha \nu_\beta \bar{\nu}_\beta) \,\approx\, \left(\frac{g^2}{M_W^2}\right)^2  
\,\frac{E^5}{192\,\pi^3} \,\left(\frac{E}{\Lambda}\right)^{3n}&\int dx' \,dx_- \,dx_+ \,\delta(1-x'-x_--x_+) \nonumber \\
& \times \frac{1}{4} \,(1 - x'^{\,n+1} - x_-^{\,n+1} + \alpha_1^{(n)} x_+^{\,n+1})^3 \,(1-x_+)^2 
\,.
\label{Gamma(spl)}
\end{align}
For the linear ($n=1$) and quadratic ($n=2$) cases, one can use the identities
\be
1 - x'^{\,2} - x_-^2 + x_+^2 \,=\, 2 \,(1-x') (1-x_-)\,, \quad\quad\quad 1 - x'^{\,3} - x_-^3 - x_+^3 \,=\, 3\,(1-x')\, (1-x_-)\,(1-x_+)\,,
\ee
to simplify the expression of the decay width,
\begin{align}
\Gamma(\nu_\alpha(E)\to \nu_\alpha \nu_\beta \bar{\nu}_\beta) \,\approx\, \left(\frac{g^2}{M_W^2}\right)^2  
\,\frac{E^5}{192\,\pi^3}  \,\left(\frac{E}{\Lambda}\right)^{3n}&\int dx' \,dx_- \,dx_+ \,\delta(1-x'-x_--x_+)\nonumber \\ & \times  \frac{(n+1)^3}{4} \,(1-x')^3 \,(1-x_-)^3 \, (1-x_+)^{3n-1}\,.
\end{align}

In the case $\beta=\alpha$ one should add a factor $1/\sqrt{2}$ in the amplitude from the antisymmetrization of the two neutrinos in the final state and one should consider two contributions related by the exchange of the momenta of the final-state neutrinos. But since the expression for the decay width is symmetric in these momenta, one concludes that the decay width for the neutrino-antineutrino pair production is flavor independent. 

After integration over the energy fractions, we obtain the following result for the total decay width producing a neutrino-antineutrino pair, valid for the $n=1,2$ cases:
\begin{empheq}[box=\boxed]{equation}
\Gamma(\nu_\alpha(E)\to \nu_\alpha \nu_\beta \bar{\nu}_\beta) \,\approx\, \left(\frac{g^2}{M_W^2}\right)^2  
\,\frac{E^5}{192\,\pi^3}  \,\left(\frac{E}{\Lambda}\right)^{3n} \,c_n^{(\nu)}\,, \quad \text{for } n=1,2 \,,
\label{Gamma-nu-nubar}
\end{empheq}
where
\be
c_n^{(\nu)} \,=\, \frac{(n+1)^3}{4} \,\left[\frac{1}{3n+1} - \frac{3}{3n+2} + \frac{7}{2(3n+3)} - \frac{2}{3n+4} + \frac{3}{5(3n+5)} - \frac{1}{10\,(3n+6)} + \frac{1}{140\,(3n+7)}\right]\,,
\ee
with numerical values given by
\be
c_1^{(\nu)} \,=\, \frac{11}{450}\approx 0.024 \,, \hskip 2cm c_2^{(\nu)} \,=\, \frac{237}{10010}\approx 0.024\,.
\label{c(nu)number}
\ee
In the same way, the energy distribution of the two neutrinos and the antineutrino, for $n=1,2$, is
\begin{empheq}[box=\boxed]{equation}
{\cal P}_{\nu_\alpha\nu_\beta\bar{\nu}_\beta/\nu_\alpha}(x', x_-, x_+) \,=\, \frac{(n+1)^3}{4\,c_n^{(\nu)}} \,\delta(1-x'-x_--x_+)\,
(1-x')^3 \,(1-x_-)^3 \, (1-x_+)^{3n-1}\,, \quad \text{for } n=1,2\,.
\label{P-nu2-nubar}
\end{empheq}

For the case $n=2$, in which both neutrinos and antineutrinos are unstable particles able to decay, one can obtain the final energy distribution in the decay of the antineutrino by replacing each particle by its antiparticle and vice versa in Eq.~\eqref{P-nu2-nubar}. The total decay width remains unchanged.

\subsection{Comparison of different decays}

The different decay widths of a superluminal neutrino can be expressed as a product of four factors
\be
\Gamma \,=\, k\, \left[E^5 \left(\frac{E}{\Lambda}\right)^{3n}\right]\, c_f\, c_n\,.
\ee
One has a common factor $k$
\be
k \,=\, \left(\frac{g^2}{M_W^2}\right)^2\, \frac{1}{192\,\pi^3} \,\approx\, 7.31\times 10^{-13} \,\text{GeV}^{-4}\,,
\ee
an energy-dependent factor \be
E^5\, \left(\frac{E}{\Lambda}\right)^{3n}\,,
\ee
 a flavor dependent factor $c_f$ due to the different couplings of the $Z$ boson to the charged leptons and the neutrinos, as well as to the additional contribution due to the $W$ exchange in the decay of $\nu_e$ producing an electron-positron pair,
\be
\begin{aligned}
c_f^{(\nu_e\to \nu_e e^- e^+)} \,=\, \left[\left(s_W^2 - \frac{3}{2}\right)^2 + (s_W^2)^2\right]\,=\,1.66 \,, \\
c_f^{(\nu_{\mu,\tau}\to \nu_{\mu,\tau} e^- e^+)} \,=\, \left[\left(s_W^2 - \frac{1}{2}\right)^2 + (s_W^2)^2\right] \,=\,0.13\,,
\end{aligned} 
\hskip 1cm
c_f^{(\nu_\alpha\to \nu_\alpha \nu_\beta \bar{\nu}_\beta)} \,=\, 1\,, 
\ee
and a factor $c_n$ [see (\ref{c(e)number}),(\ref{c(nu)number})], which depends on the produced lepton pair due to the fact that LIV only affects to the neutrinos. 
From the energy distribution (\ref{nu-nu-PP}), one has that the mean fractional energy of the daughter neutrino in the interaction producing an electron-positron pair is 
\be
\langle x \rangle \,=\, \int dx\, x\, {\cal P}_{\nu_\alpha}(x) \,=\,  \left\{ \begin{aligned}
0.26 \quad \text{for } n=1\\
0.30 \quad \text{for } n=2
\end{aligned} \right.\,.
\ee

In the case of a decay producing a neutrino-antineutrino pair, one can use the energy distribution (\ref{P-nu2-nubar}) to get the mean fractional energy of each of the two daughter neutrinos
\be
\langle x'\rangle \,=\, \langle x_-\rangle \,=\, \int dx'\, dx_-\, dx_+\, x_- \,{\cal P}_{\nu_\alpha\nu_\beta\bar{\nu}_\beta/\nu_\alpha}(x',x_-,x_+) \,=\,\left\{ \begin{aligned}
0.30 \quad \text{for } n=1\\
0.38 \quad \text{for } n=2
\end{aligned} \right.\,,  
\ee
and the mean fractional energy of the antineutrino
\be
\langle x_+\rangle \,=\, \int dx'\, dx_-\, dx_+\, x_+ \,{\cal P}_{\nu_\alpha\nu_\beta\bar{\nu}_\beta/\nu_\alpha}(x',x_-,x_+) \,=\, \left\{ \begin{aligned}
0.40 \quad \text{for } n=1\\
0.24 \quad \text{for } n=2
\end{aligned} \right. \,,
\ee
produced in the decay of the superluminal neutrino.

\section{Summary}
\label{sec:conclusions}
\medskip

We have considered the three-body decay of a superluminal particle in the very high-energy limit, where one can neglect all the masses of the particles. An explicit expression for the decay width, Eq.~\eqref{Gamma(x)}, as an integral over the energy fractions, has been obtained by an appropriate use of the collinear approximation to integrate over the angular variables. We have derived the result for a modification of the energy-momentum relation of each particle proportional to a power $n$ of the inverse of an energy scale parametrizing the correction due to LIV. As a result, we have obtained that the neutrinos are superluminal (unstable) particles for every value of $n$, but in contrast, the antineutrinos are subluminal (stable) in the linear case and superluminal in the quadratic case. If the opposite sign were chosen in \eqref{LIV-nu}, the antineutrinos would always be superluminal particles and the stability of the neutrino would depend on $n$. One can translate our study to that case just by exchanging each particle by its antiparticle in the final results.

The collinear approximation has been applied to the decay of a superluminal neutrino producing an electron-positron or a neutrino-antineutrino pair. The main results of this work are the total decay widths in \eqref{Gamma(nux-ee)}, \eqref{Gamma(nue-ee)}, \eqref{Gamma-nu-nubar}, and the energy distributions in \eqref{P-mu-tau}, \eqref{P-e}, and \eqref{P-nu2-nubar}. Compared to previous works, we have carried out a calculation which includes for the first time the charged weak current and the neutrino splitting contributions to the decay width for LIV corrections which are suppressed by a high-energy scale, as it is the case of interest in quantum gravity phenomenology.

In previous works, only the contribution mediated by the $Z$ boson was considered. In fact, the contribution mediated by the $W$ is responsible of the difference in the flavor dependent factors $c_f^{(\nu_e\to\nu_e e^- e^+)}$ and $c_f^{(\nu_{\mu,\tau}\to\nu_{\mu,\tau} e^- e^+)}$. The large value of this difference shows that neglecting the $W$ contribution is not a good approximation. 

In the case of the production of a neutrino-antineutrino pair, the lack of a calculation of the corresponding width led Ref.~\cite{Stecker:2014oxa} to approximate it by that of the production of the electron-positron pair multiplied by a factor three to take into account the three possible neutrino-antineutrino pairs. Our results show that this is not either a good approximation. 

The results of the mean fractional energies also differ from the approximation used in Ref.~\cite{Stecker:2014oxa} for the mean fractional energy $\langle x\rangle=0.22$ in the electron-positron pair production, and  $\langle x'\rangle=\langle x_-\rangle=\langle x_+\rangle=(1/3)$ in the neutrino-antineutrino pair production. A future simulation of the propagation of superluminal neutrinos could use the explicit form of the energy distributions instead of the mean fractional energies.  

Recent and near future observations of extremely high-energy neutrinos could be used to put bounds on, or identify, an effect of LIV in the propagation of astrophysical neutrinos. The study of the possible effects of LIV in the propagation of neutrinos requires the expressions for the total decay widths and neutrino energy distributions derived in this work.  

\section*{Acknowledgments}
This work is supported by Spanish grants PGC2018-095328-B-I00 (FEDER/AEI), and DGIID-DGA No. 2020-E21-17R. The work of MAR was supported by MICIU/AEI/FSE (FPI grant PRE2019-089024). This work has been partially supported by Agencia Estatal de Investigaci\'on (Spain)  under grant  PID2019-106802GB-I00/AEI/10.13039/501100011033. The authors would like to acknowledge the contribution of the COST Action CA18108 ``Quantum gravity phenomenology in the multi-messenger approach''.

\bibliography{QuGraPhenoBib}

\end{document}